*Original Article*

# Implementation of High Availability Message ISO 8583 using F5 Active-Passive Failover Method

Bahrul Ilham[1], Yanto Setiawan[2]

[1,2]Computer Science Departement, BINUS Online Learning, Bina Nusantara University, Jakarta, Indonesia.

[2]Corresponding Author : yanto.setiawan@binus.ac.id

***Abstract -*** *In this research, the system was designed to solve problems related to High Availability on FDS (Fraud Detection System) servers that cannot be loaded balanced using the Round Robin method, resulting in changes to ISO 8583 messages. As a result, a method that can be used as High availability to maintain the Availability of the FDS Server without changing the message received is required. High availability will be achieved through the Active Passive Failover method, which will transfer data flows in the event of an operational failure on the FDS server. The transfer is based on CPU Load parameters and ISO 8583 messages, which are checked at each stage. The waterfall method was used in this research. The waterfall is a straightforward classic model with a linear system flow, with the output of each stage serving as the input for the next. The primary goals of this research are to ensure that data in ISO 8583 format can be streamed without changing messages, to measure the effectiveness of the Active Passive Failover method in performing High availability using ISO 8583 message parameters and CPU Load, and to increase the level of Availability and reliability of the FDS Server. This research's error rate decreased by 0.83%, and the SLA (Service Level Agreement) increased from 99.13% to 99.96%.*

## 1. Introduction

To compete in an increasingly competitive environment, banks must continuously improve IT-based services to realize the bank's vision and mission of providing the best, excellent service and working optimally and well. Excellent service through a wide-spread network supported by quality IT systems with maximum levels of reliability and availability helps to support customer satisfaction. To address these challenges, banks require an FDS (Fraud Detection System) platform that can detect fraud in a transaction 24 hours a day, seven days a week [1].

To achieve High Availability (HA) on a bank's Critical FDS (Fraud Detection System) System, a mechanism is required to switch over when the FDS server fails to process transactions. The ISO 8583 message, the ISO message standard for financial transactions, is used for data processing on the FDS system [2]. To complete the request and response cycle of transactions originating from cards originating from ATMs, POS, or the web, financial organizations communicate with one another using ISO 8583 or its derivatives [3].

When sending ISO 8583 messages to the load-balancing FDS server, there is an issue with the message change. The issue arises when load balancing is performed using the F5 Load Balancer's iRule feature [4], which performs load balancing per message, resulting in message changes. Changes to this message prevent the intended FDS server from translating the ISO 8583 message. Adding characters and combined messages during the load-balancing process is the primary cause of this message change.

When using ISO 8583 messages, load balancing is less effective. The High Availability, Active-Passive Failover approach handles changes to this message [5]. This method will use two nodes that are both actively running the same type of service simultaneously [6]. Not all nodes will be active, as the term "active-passive" implies. In the case of two nodes, if the first is already active, the second must be passive or idle. When a node fails, the passive server (failover) serves as a backup that is ready to take over as soon as the active (main) server is disconnected or unable to serve [7]. This high availability will later use two parameters to perform the switchover [8], which are based on the ISO message and based on the value of the CPU Load from the FDS server.

Based on previous research, in 2018, there was a research journal by [9]. This research develops the architecture of High Availability on VoIP telephone systems. This research develops the architecture of High Availability on VoIP telephone systems. This study uses open-source software, Asterisk, and the SIP Protocol (Session Initiation Protocol). However, the software does not support the SIP protocol, so it lacks scalability and fault tolerance. The method is divided into several stages, namely the selection of open source software used, the topology design used, conducting test scenarios, and analyzing test results. The outcome of this research is using different tools that enable high availability and clustering concepts to offer contingency mechanisms and load balancing techniques to add greater scalability to the telephony services provided.

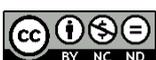

Based on the research described above, the system was created to solve problems related to High Availability on FDS (Fraud Detection System) servers that cannot load balancing using the Round Robin method, which results in changes to ISO8583 messages. High availability will be carried out by the Active Passive Failover method, which will transfer data flow when an operational failure occurs on the FDS server. The transfer is carried out based on the CPU Load parameters and ISO 8583 messages, which are checked every time. The main objectives of this research are to ensure ISO 8583 format data can be streamed without any message changes, to measure the effectiveness of the Active Passive Failover method using ISO 8583 message parameters and CPU Load in performing High availability, to measure the Availability and reliability of the FDS Server using the Active Passive Failover method. In addition, it is hoped that this research can provide benefits in the form of no changes to the ISO 8583 message to the FDS Server during the High Availability process, determine the level of Availability and reliability of the FDS Server, and as a form of risk mitigation in maintaining the reliability of the FDS Server.

The Failover technique is a network technology that provides two or more connection lines where when one of the lines dies, the connection continues by being diverted to another path [10]. This failover technique is quite important when we want a stable internet and network connection and minimal disconnection from the internet. Failover is one of the fault tolerance functions of a system that provides continuous access services. The goal is to redirect requests from system failures to system backups that can perform the first operation.

High availability is a system that can be relied upon to operate continuously without failure with high performance and operational quality [11]. The system can handle problems when problems occur, such as hardware or software updates. If a server in the Cluster fails, another server or node can immediately take over to help ensure the applications or services supported by the Cluster are kept up and running. Using a High Availability Cluster helps ensure there is no single point of failure for critical IT and reduces or eliminates downtime.

The ISO 8583 standard is a message exchange specification that defines a format for exchanging transactional data across the payment processing chain [24]. Transactions can be of various types, e.g., purchases, cash withdrawals, refunds, Etc. It also carries information regarding the merchant, his type of business, collection method, transaction amount and currency, Etc. All of this is supported through a flexible message structure. The basic message structure has three components: MTI, bitmap, and data elements.

Some research on High availability has been carried out. One of the research made use of the Distributed Replicated Block Device (DRBD) [13]. The Hadoop High Availability System is discussed in this research. This research selects the failover method by utilizing Distributed Replicated Block Device (DRBD) to store replication between failover servers and Heartbeat to detect downtime on the main server when using Linux-HA as a Hadoop High Availability solution. The method employed includes three stages: Hadoop system development, a stage before failover, and hold after failover. The results demonstrated that using Heartbeat and DRDB can reduce the failover time to less than 10 seconds. Heartbeat can handle real-time failover mirrored by DRDB in this research.

His second research focuses on implementing High availability using the Fencing method, which is the process of isolating nodes from computer clusters or protecting shared resources when a node appears to be down [14]. The issues at hand are the technical challenges of modern business, which necessitate the use of advanced security techniques and the dependability of applications and services. This research focuses on implementing high availability using open-source software to save money because it takes advantage of the operating system features used, Debian Linux. In HA, the Fencing method could work well because the complexity of the high availability test cases was limited.

The most recent research looks at the fault tolerance mechanism in the FaaS Platform (Function-as-a-Service) [15]. Active-Standby failover is the mechanism used. Fission, an open-source platform, is used as the FaaS platform. One of the primary concerns for FaaS providers is ensuring high functional availability. Indeed, commercial FaaS platforms are marketed as having high availability and built-in fault tolerance. All current FaaS platforms support a basic form of fault tolerance via function execution retries. The experimental mechanism involved comparing HA FaaS Fission AS (Active-Standby) and Fision vanilla (original version). According to the experimental results, AS outperforms vanilla in terms of response time and availability while incurring minimal overhead in resource consumption.

## 2. Methodology
The waterfall method was used in this research. The waterfall is a straightforward classic model with a linear system flow, with the output of each stage serving as the input for the next [16].

At this stage, an analysis of the requirements of the existing system is carried out, which is known that the High availability used is the F5 Load Balancer which uses a Round Robin. The problem was finding a change in the ISO 8583 message sent by the F5 Load Balancer to the FDS Server, which caused the FDS Server to crash. So the Availability of the FDS Server needs to be maintained. Then by using the Active Passive Failover Method, a needs analysis is carried out based on the available system. This method provides High availability without requiring per-message load balancing.

This stage was used to discuss a work plan, specifically by recording what tools and materials were required, both hardware and software and arranging a schedule of activities related to building this system. Furthermore, network topology design and the method used in Active Passive Failover were performed.

Implement High Availability with the Active Passive Failover Method on the FDS Server based on the simulation phase. Environment Production powers the implementation process. Implementation was only done at night or outside working hours to avoid errors during operating hours.

At this point, the researcher tested the system to reduce the error rate in the implementation. Server Environment Development was used for the trials. Simulations and trials would be carried out as close to the actual implementation as possible.

After the system has been implemented, maintenance is carried out so the system can run properly without any errors that disrupt operations. Maintenance is done by checking the process of the running system. Maintenance is carried out regularly every week to measure the effectiveness of the system's performance.

The following steps taken in designing a high-availability system can be seen in Figure 1:

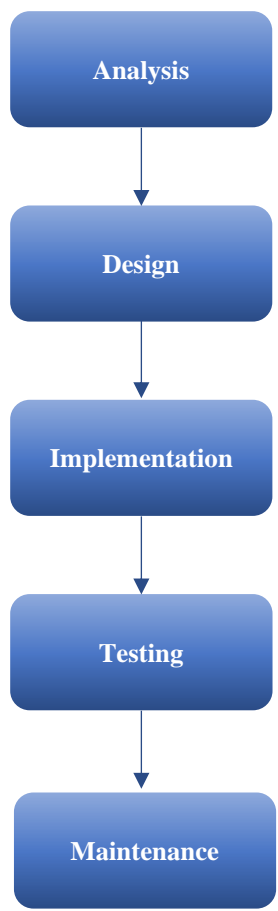

**Fig. 1 Conceptual Framework Flowchart**

## 3. Result And Discussion

A server integrated with the FDS Server and Load Balancer devices was used during the testing, failover, and stress test stages. On server testing, two FDS servers and one server load balancer with two virtual servers were used. Furthermore, Apache Jmeter was used to send messages that served as data for testing. SLA is one of the critical parameters in determining the feasibility of measuring an IT service [25]. It was hoped that this experiment would produce an SLA (Service Level Agreement) [18] in accordance with predetermined standards; additionally, it has been discovered that the number of non-standard ISO messages has decreased.

The stress test is one type of performance test. The performance test is a test that aims to verify the condition and performance of the system [19]. Stress test a two-parameter scenario, such as the amount of data to be sent and the time interval for sending data, would be used in this Stress Test experiment. Several more parameters were obtained using these two parameters, which were used as thresholds to determine the failover value. The results of the Stress Test on the FDS server are shown below. To carry out the Stress Test, three intervals were used as samples. This interval had an impact on the processing interval as well as the number of incoming transactions [20]. The number of daily transactions processed determined the amount of data received.

Based on these three results, it was clear that the error rate began to occur at 500 TPS of data with a 7%-8% increase in CPU Load. Because the average number of daily TPS data was 100 - 200 TPS, 500 TPS was a relatively high TPS value. An average daily TPS of 100-200 TPS was likely to have a low error rate. The CPU load increased by 7-8% when data errors occurred. According to the number of incoming TPS, the pattern of increase in the three Stress Tests followed the same pattern. Because of a queue for fast processing, the 1-minute interval had a high error rate increase. Because the time interval for incoming data was set to one minute, many queues went unprocessed or timed out. The 2 and 3-minute intervals took longer to process data, resulting in a shorter timeout in the data queue than the 1-minute interval. As a result, the two interval parameters and the amount of data significantly impacted the FDS Server's performance.

Figures 2 and 3 show that the trend data for the error rate tends to increase as the TPS increases. Meanwhile, CPU Load is fairly consistent, with the highest value reaching 19%. The obtained cpu Stress Test value can be used as a parameter in creating a threshold for performing a Health Check. A health check was one method of determining a server's condition [21]. When a server was running, the Health Check used parameters that would be measured as standards. This experiment will look at how Health Check worked to determine failover conditions. This experiment would use a script that would function as a parameter condition reader on the FDS Server. These were the CPU Load and ISO Message parameters. The F5 monitor would read these two parameters to determine the failover condition.

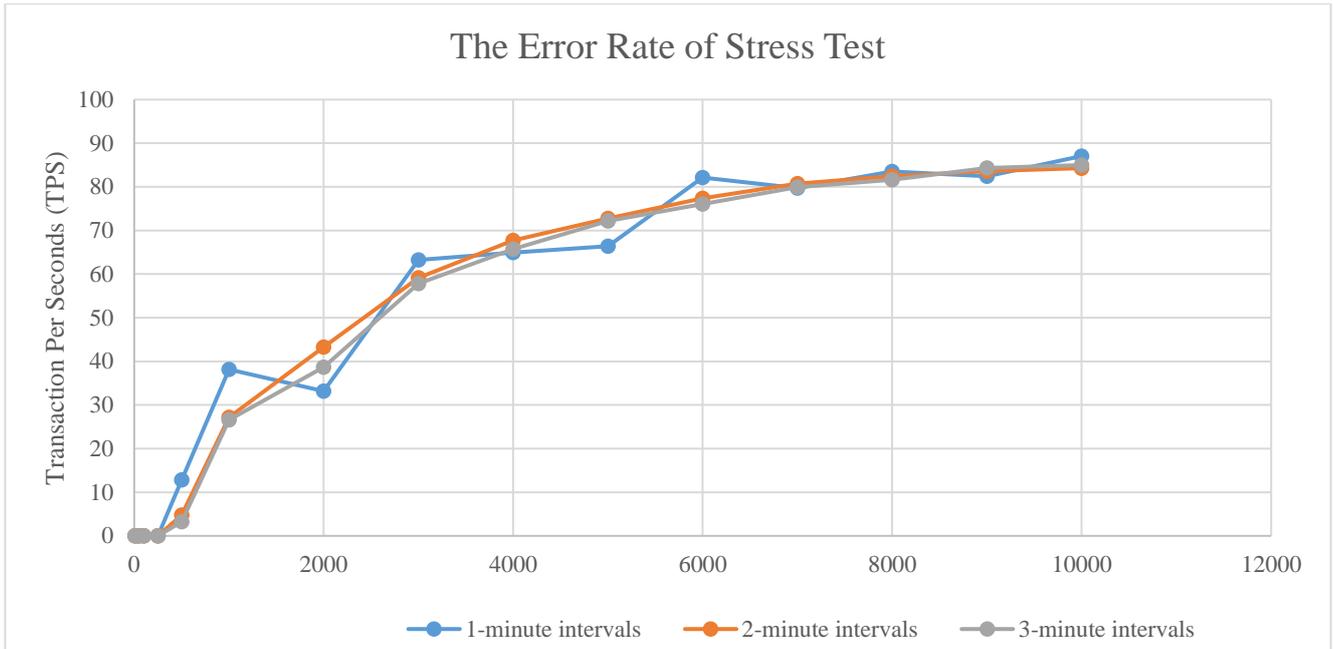

**Fig. 2 Graph of error rate on stress test**

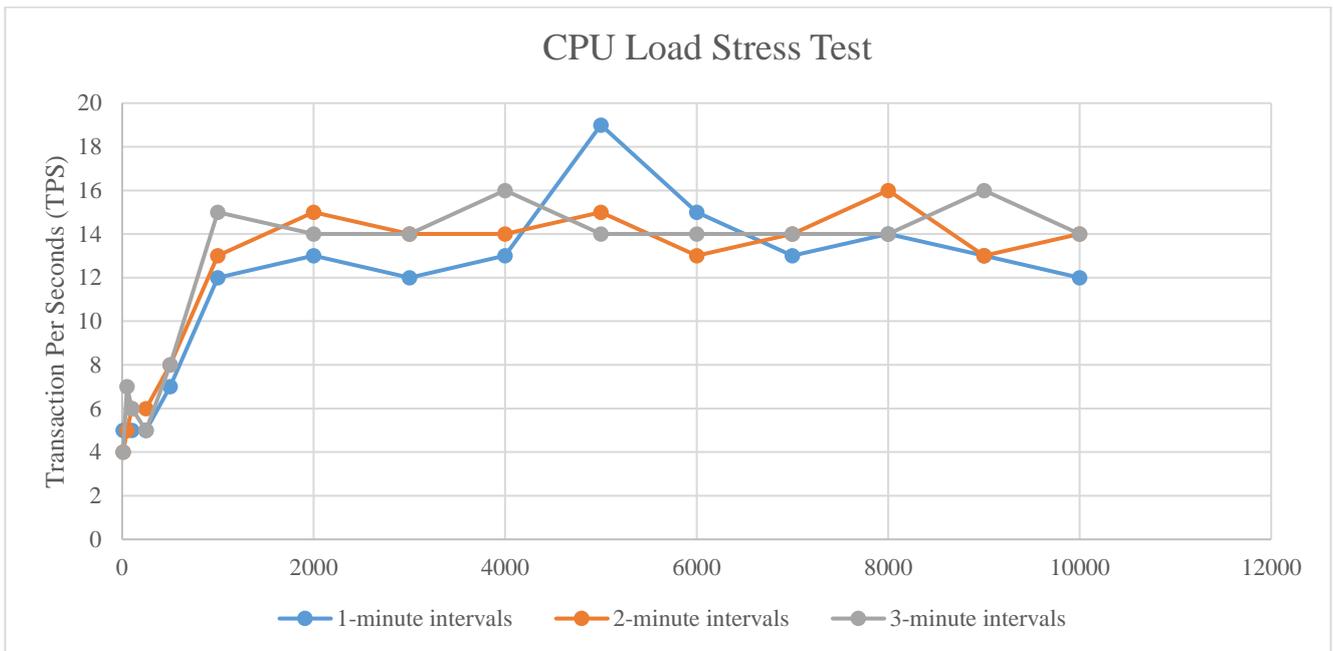

**Fig. 3 Graph of CPU Load Stress Test**

The result of the Health Check sent by the script to F5 is shown in Figure 4. The image depicts the flow of Health Check and traffic. The green status indicates that Health Check is functioning properly, and the number of data streams flowing is displayed. This Health Check would collect real-time ISO Messages and CPU Load monitoring data.

One of the techniques in High availability was the failover technique, which provided two connection lines that flow data from one connection experiencing problems to another to keep it running [22]. The availability of a server would be maintained, and it would continue to run operationally using this technique.

Figure 5 shows that the failover was completed successfully with a 20% threshold based on the results of the Stress Test experiment. A red tick in the status section of the F5 display indicated successful failover. Data streams that previously flowed to both servers, namely to 172.18.129.103 and 172.18.132.116 on port 9050, only flew to 172.18.129.103 due to an increase in CPU load on Server 172.18.132.116 with a 20% threshold.

Figure 6 shows that the ISO Message-based failover was successful, as indicated by the red status on the F5 display. Because no ISO messages were processed on the FDS Server because the processing service was not running normally or was down, the flow on the FDS Server with IP 172.18.132.116 was successfully interrupted.

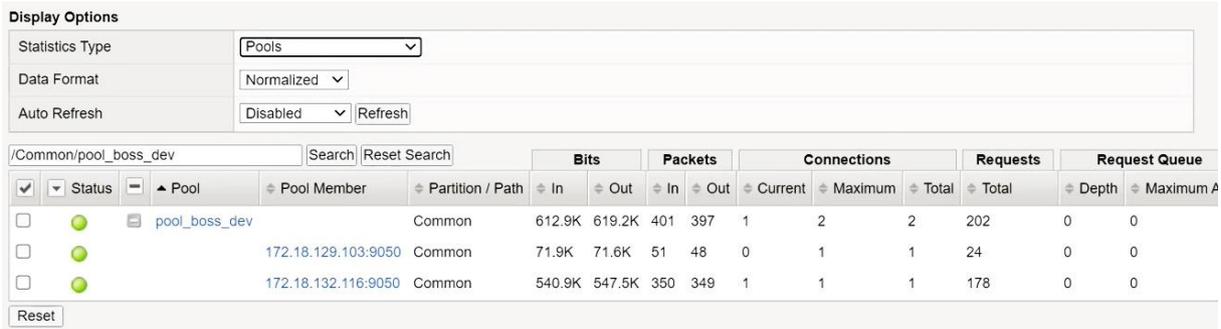

Fig. 4 Health Check and Data Traffic F5

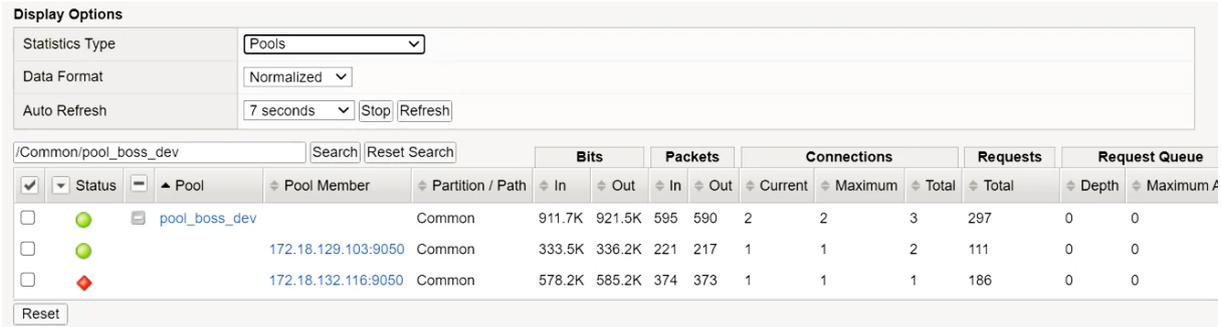

Fig. 5 Condition of Failover CPU Load

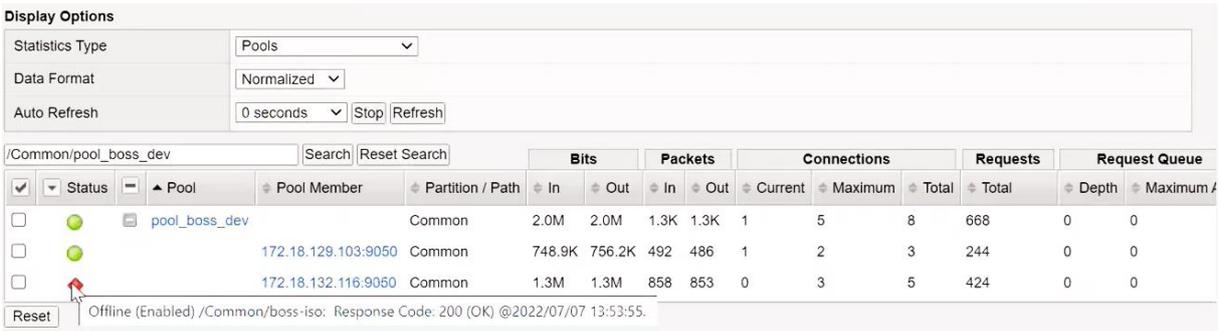

Fig. 6 Condition of Failover ISO Message

ISO 8583 formatted messages are an alphanumeric data set arranged according to specific rules and read according to certain rules as well according to the rules of ISO 8583 [24]. This research used a monitoring scenario on the implementation results of failover in this experiment. The number of non-standard messages would have decreased or increased as a result of this implementation [23]. The results of scalability experiments on ISO message changes on the FDS Server are as follows.

The results of monitoring the flow of data to the FDS Server in April before applying the failover method can be seen in Figure 7. The data were collected daily by counting the number of non-standard ISO messages, standard ISO messages, and the total number of incoming transactions. The percentage of each parameter was also calculated using these data. This monitoring data yields a mean of the number of daily transactions was 4,036,093 transactions; The number of daily standards ISO Messages was 4,000,926 transactions; The number of non-standard ISO messages per day was 35,167 transactions; Then the daily TPS value was 4,036,093 transactions / 86400 seconds = 46.71 TPS; Total daily error rate was 0.867%, and SLA (Service Level Agreement) was 99.13%.

Data collected every day in October can be seen in Figure 8. From the data obtained, the percentage data of each parameter was. Here are the averages from the table: The number of daily transactions was 3,158,269 transactions; The number of daily standards ISO Messages was 3,157,096 transactions; The number of non-standard ISO messages per day was 1,173 transactions; Then the daily TPS value was 3,158,269 transactions / 86400 seconds = 36.55 TPS; Total daily error rate was 0.037%, and SLA (Service Level Agreement) was 99.96%. Data following failover implementation showed an average daily error rate of 0.037% and an SLA of 99.96%. The error rate has dropped from 0.867% to 0.037%. As a result, the error rate has decreased by 0.83%.

Furthermore, the SLA for successfully processed transactions was increased from 99.13% to 99.96%. There was a 0.83% increase. This indicated improved transaction processing performance and a decrease in the number of non-standard ISO messages.

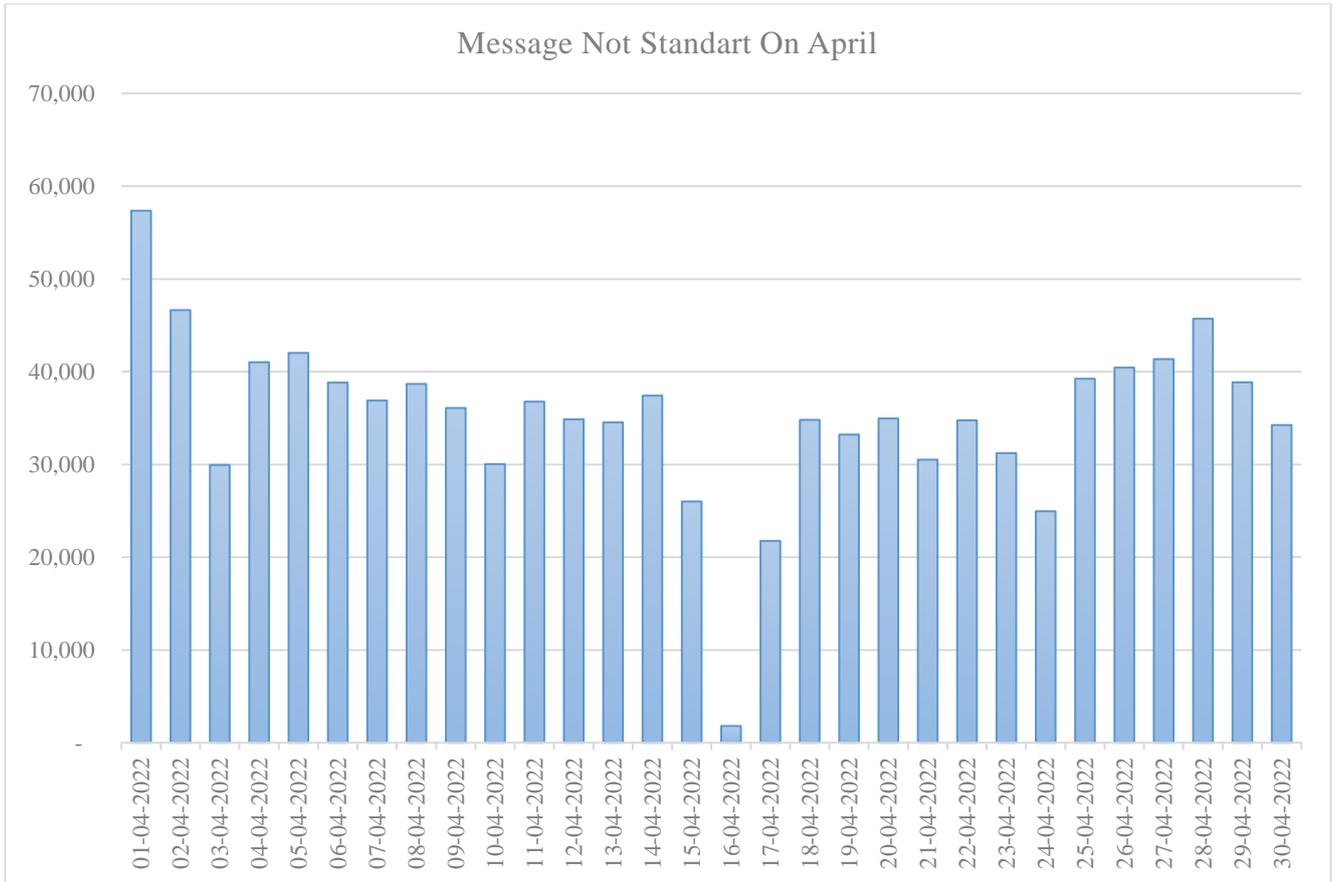

**Fig. 7 Graph of Message Not Standart on April**

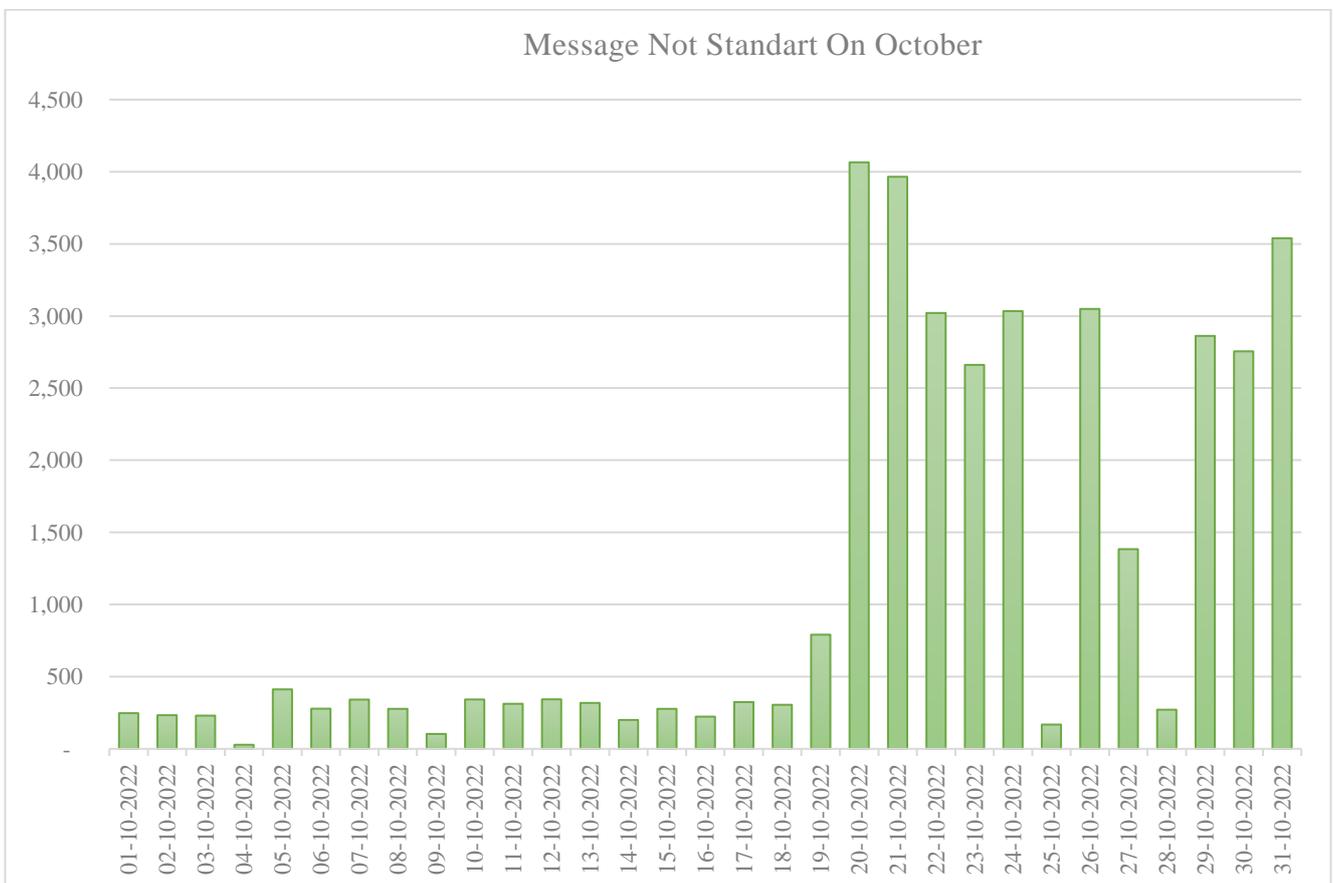

**Fig. 8 Graph of Message Not Standart on October**

The resulting error rate and CPU load value were generated in the Stress Test that was performed using the interval and TPS (Transaction Per Second) variables. The experiment provided an overview of the dependability of the server that would be used to process some transactions, allowing for appropriate capacity planning to accommodate the volume of incoming transactions. The results of this Stress Test were used to establish Health Check parameters. In this Stress Test experiment, a two-parameter scenario will be used, such as the amount of data sent and the time interval for sending data. By using these two parameters, several more parameters will be obtained, which are used as thresholds in determining the failover value. The following results of the Stress Test on the FDS server can be seen in Tables 1, 2, and 3.

From these three results, it can be seen that the error rate starts to occur at the amount of data of 500 TPS with an increase in CPU Load between 7% -8%. The number of 500 TPS is a fairly high TPS value because the average daily TPS data is 100 – 200 TPS. A small error rate will likely occur in an average daily TPS of 100-200 TPS. As for the CPU Load, there is an increase of 7-8% when data errors start. The pattern of increase in the three Stress Tests occurred similarly according to the number of incoming TPS. The 1-minute interval has a high error rate increase due to a queue for fast processing. The time interval for incoming data is 1 minute, so many queues are not processed or timeout.

In comparison, the 2 and 3-minute intervals have a longer time processing data, so the timeout in the data queue is smaller than the 1-minute interval. So that the 2 interval parameters and the amount of data greatly affect the performance of the FDS Server. The smaller and higher amount of incoming data will cause many data queues to be processed.

Health Check was used to provide information on the status of a running server based on parameters set to threshold values and conditions. This health check would provide a real-time value for the CPU Load parameter and the ISO Message condition, which would be used to determine the failover condition later on. The script sent the results of the Health Check to F5. Figure 9 shows that the Health Check and traffic are smooth. The green status indicates that Health Check is running well, accompanied by the number of data streams flowing. This Health Check will run in real-time to get ISO Messages and CPU Load monitoring results.

Failover assisted the server in providing backup data flow routes for transactions when the main server failed. The failover method used was a switch from the primary FDS Server to the backup FDS Server. The flow route in the failover results changed automatically based on the monitoring results from the health check.

Table 1. 1-Minute Interval Stress Test Experiment

| Amoun data (TPS) | Interval | Samples | Error Rate (%) | CPU (%) |
|---|---|---|---|---|
| 10 | 1 minute | 434 | 0 | 5 |
| 50 | 1 minute | 860 | 0 | 5 |
| 100 | 1 minute | 908 | 0 | 5 |
| 250 | 1 minute | 1005 | 0 | 5 |
| 500 | 1 minute | 1552 | 12.82 | 7 |
| 1000 | 1 minute | 3173 | 38.12 | 12 |
| 2000 | 1 minute | 3663 | 33.2 | 13 |
| 3000 | 1 minute | 6547 | 63.22 | 12 |
| 4000 | 1 minute | 7805 | 64.95 | 13 |
| 5000 | 1 minute | 9644 | 66.39 | 19 |
| 6000 | 1 minute | 11833 | 82.13 | 15 |
| 7000 | 1 minute | 13742 | 79.72 | 13 |
| 8000 | 1 minute | 16106 | 83.52 | 14 |
| 9000 | 1 minute | 15616 | 82.42 | 13 |
| 10000 | 1 minute | 20590 | 87.06 | 12 |

**Table 2. 2-Minute Interval Stress Test Experiment**

| Amoun data (TPS) | Interval | Samples | Error Rate (%) | CPU (%) |
|---|---|---|---|---|
| 10 | 2 minute | 860 | 0 | 5 |
| 50 | 2 minute | 1700 | 0 | 5 |
| 100 | 2 minute | 1740 | 0 | 5 |
| 250 | 2 minute | 1853 | 0 | 5 |
| 500 | 2 minute | 3180 | 4.81 | 7 |
| 1000 | 2 minute | 4995 | 27.15 | 12 |
| 2000 | 2 minute | 7190 | 43.25 | 13 |
| 3000 | 2 minute | 10067 | 59.16 | 12 |
| 4000 | 2 minute | 13148 | 67.69 | 13 |
| 5000 | 2 minute | 15648 | 72.74 | 19 |
| 6000 | 2 minute | 19045 | 77.38 | 15 |
| 7000 | 2 minute | 22095 | 80.72 | 13 |
| 8000 | 2 minute | 24731 | 82.44 | 14 |
| 9000 | 2 minute | 26020 | 83.57 | 13 |
| 10000 | 2 minute | 28107 | 84.29 | 12 |

**Table 3. 3-Minute Interval Stress Test Experiment**

| Amoun data (TPS) | Interval | Samples | Error Rate (%) | CPU (%) |
|---|---|---|---|---|
| 10 | 3 minute | 1310 | 0 | 5 |
| 50 | 3 minute | 2540 | 0 | 5 |
| 100 | 3 minute | 2581 | 0 | 5 |
| 250 | 3 minute | 2740 | 0 | 5 |
| 500 | 3 minute | 4980 | 4.81 | 7 |
| 1000 | 3 minute | 6809 | 27.15 | 12 |
| 2000 | 3 minute | 9606 | 43.25 | 13 |
| 3000 | 3 minute | 13661 | 59.16 | 12 |
| 4000 | 3 minute | 17607 | 67.69 | 13 |
| 5000 | 3 minute | 21608 | 72.74 | 19 |
| 6000 | 3 minute | 25327 | 77.38 | 15 |
| 7000 | 3 minute | 29958 | 80.72 | 13 |
| 8000 | 3 minute | 33632 | 82.44 | 14 |
| 9000 | 3 minute | 37667 | 83.57 | 13 |
| 10000 | 3 minute | 41336 | 84.29 | 12 |

**Fig. 9 Result Script Health Check**

## 4. Conclusion
After testing the failover method on the FDS Server, it is possible to conclude that implementing high availability on the FDS Server requires several stages. In the first stage, a stress test is required as a reference material in developing a threshold that will be used in monitoring health checks. This monitoring health check informs the system about the state of the CPU load and the ISO Message. By conducting a stress test, an overview of the reliability of the FDS Server is obtained. In addition, with the stress test, the maximum value of CPU Load is obtained, which is used as a health check parameter. Implementing a health check will help the failover process in monitoring and providing threshold values for CPU load and the condition of the absence of incoming ISO messages. The failover system helps switch the transaction flow from the primary FDS Server to the backup FDS Server when a failure occurs on the primary server to maintain high availability on the FDS Server. With the failover method, this high availability system reduces the error rate to the number of non-standard ISO messages from 0.867% to 0.037% of the total number of incoming transactions. 5. The high availability system with the failover method increases the Service Level Agreement (SLA) value by 0.83% of the number of incoming transactions. Based on the conclusions, here are some suggestions for future research development: 1. Using a stress test as a form of capacity planning for the number of FDS Servers to be used. 2. Conduct a stress test with more interval variables for a more accurate health check threshold value. 3. Increasing the number of FDS Servers used so that there are more options for failover.